\documentclass[onecolumn]{aastex62}
\usepackage{textcomp}
\usepackage{amssymb}

\hypersetup{linkcolor=red,citecolor=cyan,filecolor=blue,urlcolor=magenta}
\usepackage{subfigure}

\begin{document}

\title{Awakening of two $\gamma$-ray high redshift flat-spectrum radio quasars in the southern hemisphere}

\correspondingauthor{Neng-Hui Liao,Yi-Zhong Fan}
\email{nhliao@gzu.edu.cn, yzfan@pmo.ac.cn}

\author{Shang Li}
\affiliation{Key Laboratory of Dark Matter and Space Astronomy, Purple Mountain Observatory, Chinese Academy of Sciences, Nanjing 210034, China}

\author{Lu-Ming Sun}
\affiliation{Key laboratory for Research in Galaxies and Cosmology, Department of Astronomy, University of Science and Technology of China, Hefei, Anhui 230026, China}

\author{Neng-Hui Liao}
\affiliation{Department of Physics and Astronomy, College of Physics, Guizhou University, Guiyang 550025, China}

\author{Yi-Zhong Fan}
\affiliation{Key Laboratory of Dark Matter and Space Astronomy, Purple Mountain Observatory, Chinese Academy of Sciences, Nanjing 210034, China}

\begin{abstract}

High-redshift blazars are valuable tools to study the early Universe. So far only a handful of $\gamma$-ray blazars have been found at redshifts above 3. $\gamma$-ray signals are detected in the direction of PMN J2219-2719 ($z=3.63$) and PMN J2321-0827 ($z=3.16$) by analyzing the 10-year $Fermi$-LAT Pass 8 data. PMN J2219-2719 is not distinguished from the background in the global analysis. During the 5-month epoch, the TS value is 47.8 and the flux is more than 10 times of  the 10-year averaged flux. In addition, the angular distance between the $\gamma$-ray position and the radio position of PMN J2219-2719 is only ${0.04}^{\circ}$. Moreover, the $\gamma$-ray and infrared light curves of long time scale are very similar, which support the association between the $\gamma$-ray source and PMN J2219-2719. The global analysis of PMN J2321-0827 suggest a new $\gamma$-ray source, during the flare phase, the TS value is 61.4 and the $\gamma$-ray flux increased significantly. The association probability suggests that PMN J2321-0827 may be the counterpart of the new $\gamma$-ray source. In the future, the number of high-redshift $\gamma$-ray sources will increase by combining $Fermi$-LAT and the upcoming Large Synoptic Survey Telescope.

\end{abstract}


\section{Introduction} \label{sec:intro}
Blazars, known as a peculiar subclass of active galactic nuclei (AGN) including flat-spectrum radio quasars (FSRQs) and BL Lacertae objects (BL Lacs), are among the most energetic phenomena in the universe \citep[e.g.,][]{2014Natur.515..376G}. Their jets are well aligned with our line of sight and hence the jet emissions are strongly boosted due to the relativistic beaming effects \citep{1978bllo.conf..328B}. The non-thermal jet emission is characterized by violent variability with variation timescales spanning from minutes to years \citep[e.g.,][]{1997ARA&A..35..445U}. Meanwhile, it shows a universal two-bump structure in log$\nu$F$\nu$-log$\nu$ plot. The first bump is attributed to be synchrotron emission while the other one extends to $\gamma$-ray domain. Blazars are the dominant population in the extragalactic $\gamma$-ray sky. There are about three thousand sources in 3FGL catalog \citep{2015ApJS..218...23A} and nearly five thousands in 4FGL catalog (4FGL, \citealt{4FGL}). The number of sources in 4FGL, nearly three thousand are AGNs or blazars (4FGL, \citealt{4FGL}). In the leptonic scenario, the $\gamma$-ray emission of blazars is usually explained as inverse Compton (IC) scattering of soft photons from either inside (the synchrotron self-Compton, or SSC) and/or outside of the jet (external Compton, or EC) by the same population of relativistic electrons that give rise to the synchrotron emission \citep[e.g.,][]{1981ApJ...243..700K,1985ApJ...298..114M,1994ApJ...421..153S,2000ApJ...545..107B}. Alternatively, the coincidence between the arrival of the PeV neutrino and the $\gamma$-ray flares in TXS 0506+056 indicates that at least in some cases hadronic processes could play an important role \citep{2018Sci...361.1378I}.

Within the entire blazar population, high-redshift blazars are of special interest \citep[e.g.,][]{2004ApJ...610L...9R,2006AJ....132.1959R}. On one hand, they are valuable targets for understanding the formation and growth of the first generation of super massive black holes (SMBHs) as well as the cosmic evolution of AGN jets \citep{2010MNRAS.405..387G,2010A&ARv..18..279V}. On the other hand, their emissions carry crucial information of the early universe. Particularly, the $\gamma$-ray emission of high-redshift blazars is a powerful tool for probing the extragalactic background light (EBL, e.g., \citealt{2018Sci...362.1031F}). Stacking analysis suggests that high-redshift blazars are faint $\gamma$-ray emitters \citep{2020arXiv200601857P}. In consideration of their large distances, it is not surprising that the number of detected high redshift (i.e. $z >$ 3) $\gamma$-ray blazars is rather limited. Since so far all blazars detected by {\it Fermi}-LAT beyond redshift 3 are FSRQs \citep{4FGL} and the peak of their high energy SED bump is beneath the lower energy threshold of {\it Fermi}-LAT, the significant cosmic redshift makes detection of $\gamma$-ray emission of high-redshift blazars more challenging. Right now there are only handful of $\gamma$-ray blazars with redshifts over 3. In addition to the two bright blazars ($z\simeq$ 3) included in 3FGL \citep{2015ApJ...810...14A}, detections of five new such sources have been reported by {\it Fermi}-LAT Collaboration \citep{2017ApJ...837L...5A}, with the most distant one (i.e. NVSS J151002+570243) at the redshift of 4.3. The 4FGL \citep{4FGL} has embraced four additional sources\footnote{However, two sources reported in \cite{2017ApJ...837L...5A}, NVSS J064632+445116 ($z$ = 3.4) and NVSS J212912$-$153841 ($z$ = 3.3), are absent in this latest catalogue\citep{4FGL}.}. Interestingly, a spectrally soft transient $\gamma$-ray source towards B3 1428+422 ($z=4.7$), which is the farthest source among the 105 month {\it Swift}-BAT all sky hard X-ray survey \citep{2018ApJS..235....4O}, has been identified \citep{2018ApJ...865L..17L}. Currently, this source holds the redshift record for the GeV $\gamma$-ray emitter candidate.

PMN J2219-2719 and PMN J2321-0827 \citep{1994ApJS...90..179G,Braude1981} have been realized as flat-spectrum radio sources \citep{1978AJ.....83.1036C,2002A&A...386...97J,2007ApJS..171...61H}. Based on the NRAO VLA Sky Survey (NVSS, \citealt{1998AJ....115.1693C}) as well as the SuperCOSMOS data \citep{2011MNRAS.417.2651M}, their radio loudness (RL = $\rm f_{1.4~GHz}/f_{B~band}$, \citealt{1989AJ.....98.1195K}) values can be estimated as high as $\sim$ 20000 and $\sim$ 30000, respectively. X-ray observations by {\it Chandra} at different epochs reveal significant X-ray variability of PMN J2219-2719 \citep{2006AJ....131.1914L,2011ApJ...738...53S}. In addition, a hard X-ray spectrum indicative of overwhelming jet contribution has been also detected by {\it Chandra} \citep{2011ApJ...738...53S}. Meanwhile, a comparison between the SDSS ($r^{sdss}_{mag} = 20.96$, \citealt{2008ApJS..175..297A}) and SuperCOSMOS ($R_{mag} = 20.08$, \citealt{2011MNRAS.417.2651M}) data of PMN J2321-0827 suggests the existence of significant optical variability. Therefore, it is reasonable that they are embraced in the Roma-BZCat (hereafter BZCAT, \citealt{2009A&A...495..691M}) and categorized as BZQ J2219-2719 and BZQ J2321-0827, respectively. The redshift estimations, PMN J2219-2719 ($z$ = 3.63, \citealt{2002A&A...391..509H}) and PMN J2321-0827 ($z$ = 3.169, \citealt{Titov2011}), are given by spectroscopic observations, from which Ly$\alpha$ and $\rm C_{IV}$ emission lines are distinct. In this work, we report on analyses of $Fermi$-LAT $\gamma$-ray data of PMN J2219-2719 and PMN J2321-0827, aiming to increase the number of $\gamma$-ray blazars beyond redshift 3. In section 2, the detailed consideration in our $Fermi$-LAT data analysis procedure is introduced; results of the analysis are reported in section 3; discussions are presented in section 4; finally, we summarized our results. In the following paper, we take a $\Lambda$CDM cosmology with $H_{0}=67~{\rm km~ s^{-1}~Mpc^{-1}}$, $\Omega_{\rm m}=0.32$, and $\Omega_{\Lambda}=0.68$ \citep{2014A&A...571A..16P}.

\section{Data Analysis} \label{sec:data}
\subsection{{\it Fermi}-LAT data analysis}
Here the $\gamma$-ray data (i.e. $Fermi$-LAT Pass 8 data, {\tt P8R3\_SOURCE\_V2}, {\tt FRONT+BACK}) analyses, performed by {\tt Fermitools} software version 1.2.23, are based on the first ten-year (i.e. from 2008 August 4 to 2018 August 5) $Fermi$-LAT survey. The energy range of the data set is between 100 MeV and 500 GeV. In order to suppress the contamination from the Earth's limb, the $\gamma$-ray events with zenith angle greater than $90^{\circ}$ are eliminated. Meanwhile, the recommended quality-filter cuts ({\tt DATA\_QUAL==1 \&\& LAT\_CONFIG==1}) are also applied.  We use the {\tt Unbinned} likelihood analyses method in the {\tt gtlike} task to derive the $\gamma$-ray flux and spectrum, in which photons within $10^{\circ}$ regions of interest (ROIs) centered at locations of the targets are  focused. The initial background models are generated by the script {\tt make4FGLxml.py}\footnote{\url{https://fermi.gsfc.nasa.gov/ssc/data/analysis/user/make4FGLxml.py}}, including all 4FGL\footnote{\url{https://fermi.gsfc.nasa.gov/ssc/data/access/lat/8yr_catalog/gll_psc_v21.fit}} sources within 15$^{\circ}$ around the targets as well as the diffuse $\gamma$-ray emission templates (i.e. {\tt gll\_iem\_v07.fits} and {\tt iso\_P8R3\_SOURCE\_V2\_v1.txt}). Since our targets have not been included in any current $\gamma$-ray catalogs, $\gamma$-ray sources located at the radio positions of the two high redshift FSRQs, with an assumption of single power-law (i.e. $dN/dE \propto E^{-\Gamma}$, $\Gamma$ is the spectral photon index) spectrum model, are added into the analysis model file. During the likelihood analysis, parameters of the targets and the 4FGL background sources lying within the ROIs, as well as the normalizations of the two diffuse emission backgrounds are set free. 
The test statistic (TS, \citealt{1996ApJ...461..396M}) is used to quantify the significance of a $\gamma$-ray source, which is expressed as $TS=-2{\rm ln}({L_{0}/L}$) where ${L}$ and $L_{0}$ are the best fit likelihood values for the model with and without the putative target source, respectively. When the TS value is less than 10, we use the {\tt pyLikelihood  UpperLimits} tool to calculate the 95\% confidential level (C.L.) upper limit instead of estimating the flux. Furthermore, $\gamma$-ray light curves are also extracted, in which we fixed the spectral parameters of background sources with the values from the global fit, unless they are brighter than or comparable to the target. Meanwhile, the weak (TS $<$ 10) background sources are removed from the model during the temporal analysis.

\subsection{{\it WISE} data analysis}

The Wide field Infrared Survey Explorer\footnote{\url{https://www.nasa.gov/mission_pages/WISE/main/index.html}} \citep[WISE,][]{2010AJ....140.1868W} telescope has been conducting a repetitive all-sky survey since 2010, except for a gap between 2011 and 2013. For a typical sky location, the WISE telescope visits it every half a year, and takes $>$10 exposures during $\sim$1 days. Although initially 4 filters were used, most of the time only two filters, named W1 and W2, were used. The central wavelengths of the two filters are 3.4 and 4.6 $\mu$m, corresponding to rest-frame wavelengths of 0.7 and 1.0 $\mu$m, respectively. For some sources are relatively faint and marginally detected in any single-exposure image. Thus we started from time-resolved coadds which were generated by \cite{2018AJ....156...69M} by stacking the single-exposure images taken during each WISE's visit. We performed PSF-fitting photometry on each coadd following \cite{2014arXiv1410.7397L} and during the fitting the position was fixed to that from infrared survey. In this way we obtained light curves sampled once half a year from 2010 to 2019 at rest-frame 0.7 and 1.0 $\mu$m.

For these two sources, only PMN J2219-2719 is included in AllWISE Source Catalog. Therefore we only explored whether there was corresponding infrared variation in PMN J2219-2719 during the time when the $\gamma$-ray flare was detected, using the data from WISE. PMN J2219-2719 was detected during every WISE's visit. It's infrared flux was steady for most of the time from 2010 to 2019 and showed a rise during the two WISE's visits in November 2015 and May 2016. We referred to the time period showing flux rise as a high state, and the rest of time as a low state. In the low state, the flux of PMN J2219-2719 at 0.7$\mu$m was around 65 $\mu$Jy with a standard deviation of 12 $\mu$Jy, and the flux at 1.0 $\mu$m was around 87 $\mu$Jy with a standard deviation of 24 $\mu$Jy. In the high state, the flux at 0.7 $\mu$m was $133\pm9$ and $112\pm8$ $\mu$Jy (at MJD 57575 and MJD 57698, respectively), which is 2.0 and 1.7 times the mean flux in the low state. The significance of the flux enhancements are at the level of 4.4 and 3.1 $\sigma$, respectively, and when calculating the significance, both the measurement error in the high state and the internal dispersion in the low state are considered. The flux observed at 1.0 $\mu$m was $169\pm19$ and $147\pm18$ $\mu$Jy (at MJD 57575 and MJD 57698, respectively), which is 1.9 and 1.7 times the mean flux in the low state, and the significance of the flux increase is 2.7 and 2.0 $\sigma$. These indicate that there was an infrared flare in the PMN J2219-2719 between 2015 and 2016, during which the infrared flux increased significantly (see the lower panel of Figure \ref{fig:subfigure21}).

\section{Results} \label{sec:resul}
\subsection{$\gamma$-ray properties}
Since we analyze the first 10-year $Fermi$-LAT data while the 4FGL sources are based on the first 8-year survey, we check whether there are new $\gamma$-ray sources (TS $>$ 25) arising. The residual TS map of these two sources are generated from fits of the entire dataset and there are indeed several new background sources. Majority of them are not close to the targets ($>$ 3$^\circ$) and faint (TS $<$ 100), except a bright $\gamma$-ray flare in July 2016 likely associating with a flat-spectrum radio source PKS 2247-131 (\citealt{2007ApJS..171...61H}; which is about 9$^\circ$ from PMN J2321-0827) that has been reported in \cite{2016ATel.9285....1B}. After updating the background model files, the likelihood analyses are carried out again. No significant $\gamma$-ray excess (TS $\sim$ 5) has been found towards PMN J2219-2719 (see Figure \ref{fig:subfigure11}). On the other hand, there is a $\gamma$-ray source (TS $\sim$ 27) in the direction of PMN J2321-0827 \footnote{When preparing for this manuscript, we noted that the $Fermi$-LAT collaboration presented a new version of the fourth $Fermi$-LAT catalog of gamma-ray sources (4FGL-DR2, \citealt{Ballet:2020hze}), in which PMN J2321-0827 was included} (see Figure \ref{fig:subfigure31}). The optimized position are R.A. 350.15$^\circ$ and decl. -8.38$^\circ$, with 95\% C.L. radii of 0.22$^\circ$. The radio position of PMN J2321-0827 is offset by 0.19$^\circ$ from the $\gamma$-ray location. Then, we derived a half-year bin $\gamma$-ray light curves for the two sources (see the middle panel of Figure \ref{fig:subfigure21} and the upper panel of Figure \ref{fig:subfigure41}). In the following, we present more details of the analysis.

\subsubsection{PMN J2219-2719}

\begin{figure}
\centering
\subfigure[]{%
  \includegraphics[width=0.32\textwidth]{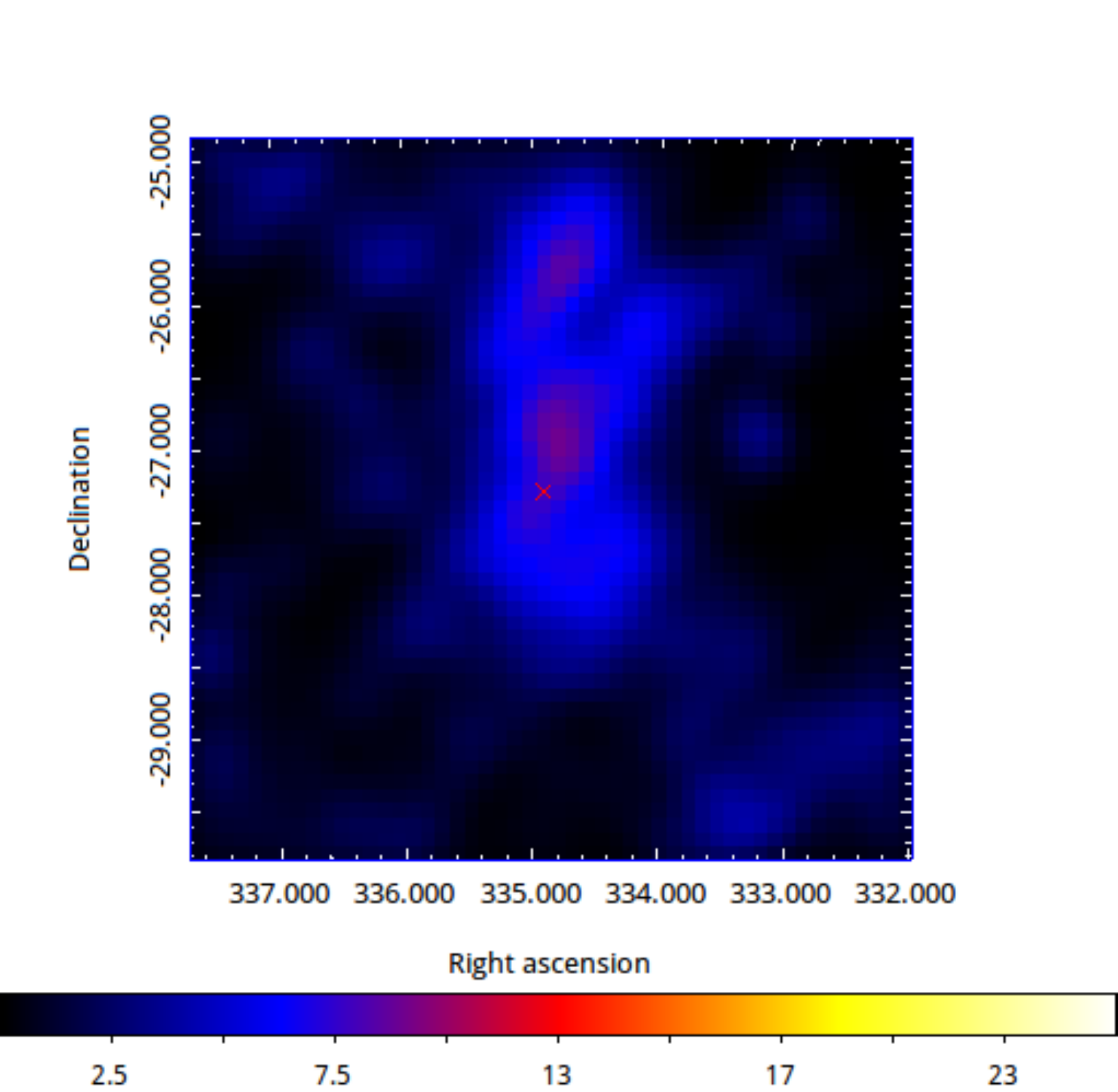}
  \label{fig:subfigure11}}
\subfigure[]{%
  \includegraphics[width=0.32\textwidth]{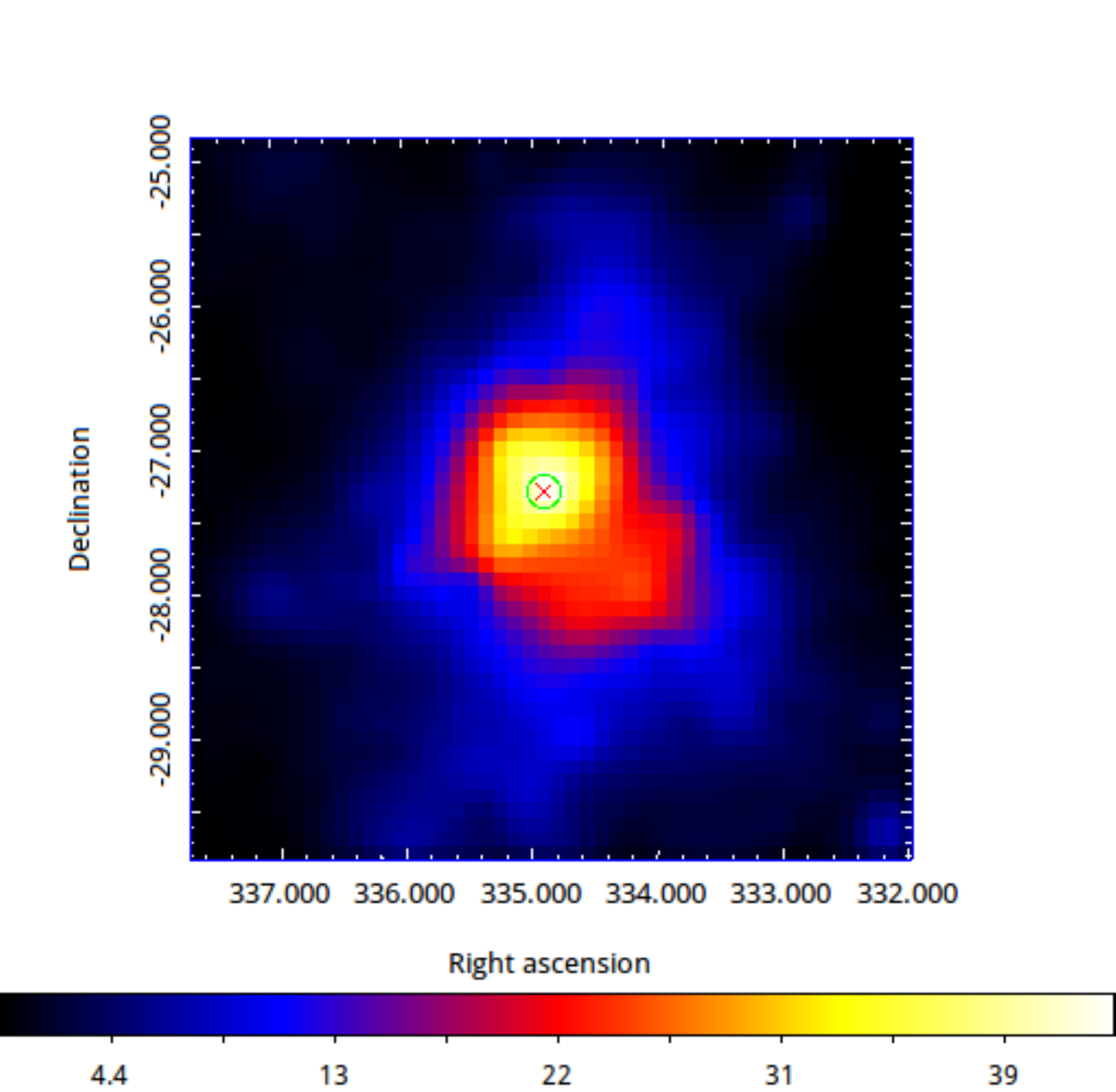}
  \label{fig:subfigure12}}
\subfigure[]{%
  \includegraphics[width=0.32\textwidth]{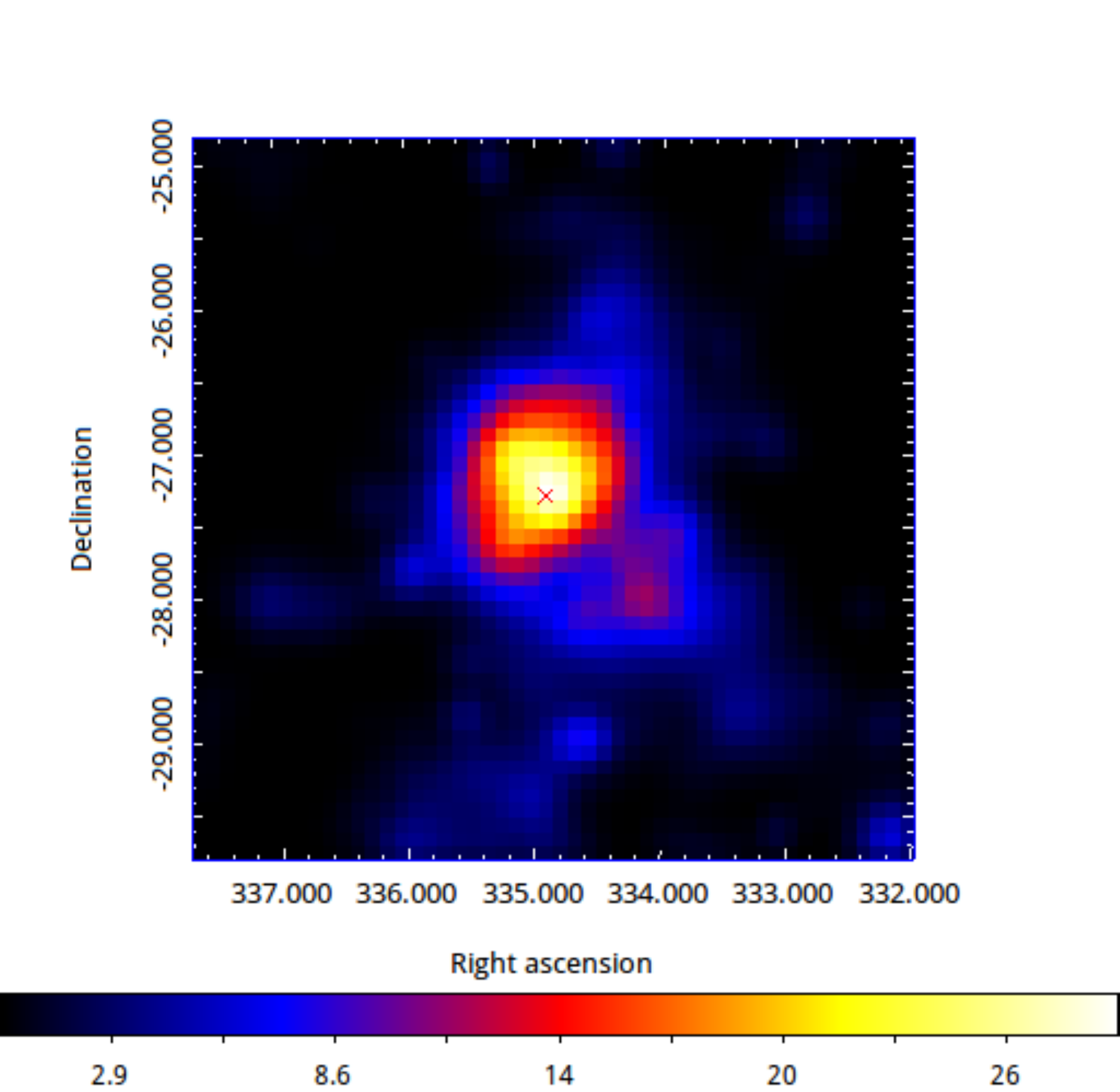}
  \label{fig:subfigure13}}
  \caption{TS maps for different epochs of PMN J2219-2719. The target source is not included in the model file. The three panels represent $5^{\circ} \times 5^{\circ}$ TS maps with $0.1^{\circ}$ per pixel centered at PMN J2219-2719. (a) The 10-year TS map between 0.1 and 500 GeV. (b) 5-month flaring epoch TS map between 0.3 and 500 GeV. (c) TS map for the flaring epoch between 0.3 and 500 GeV. The radio position (J2000) is marked in X-shaped symbol. The green circle in panel (b) is the 95\% error circle of the $\gamma$-ray position. Locations of the nearby background sources considered in the analysis are marked.}
\label{Fig.1}
\end{figure}

\begin{figure*}  
\centering
\subfigure[]{%
  \includegraphics[width=0.41\textwidth]{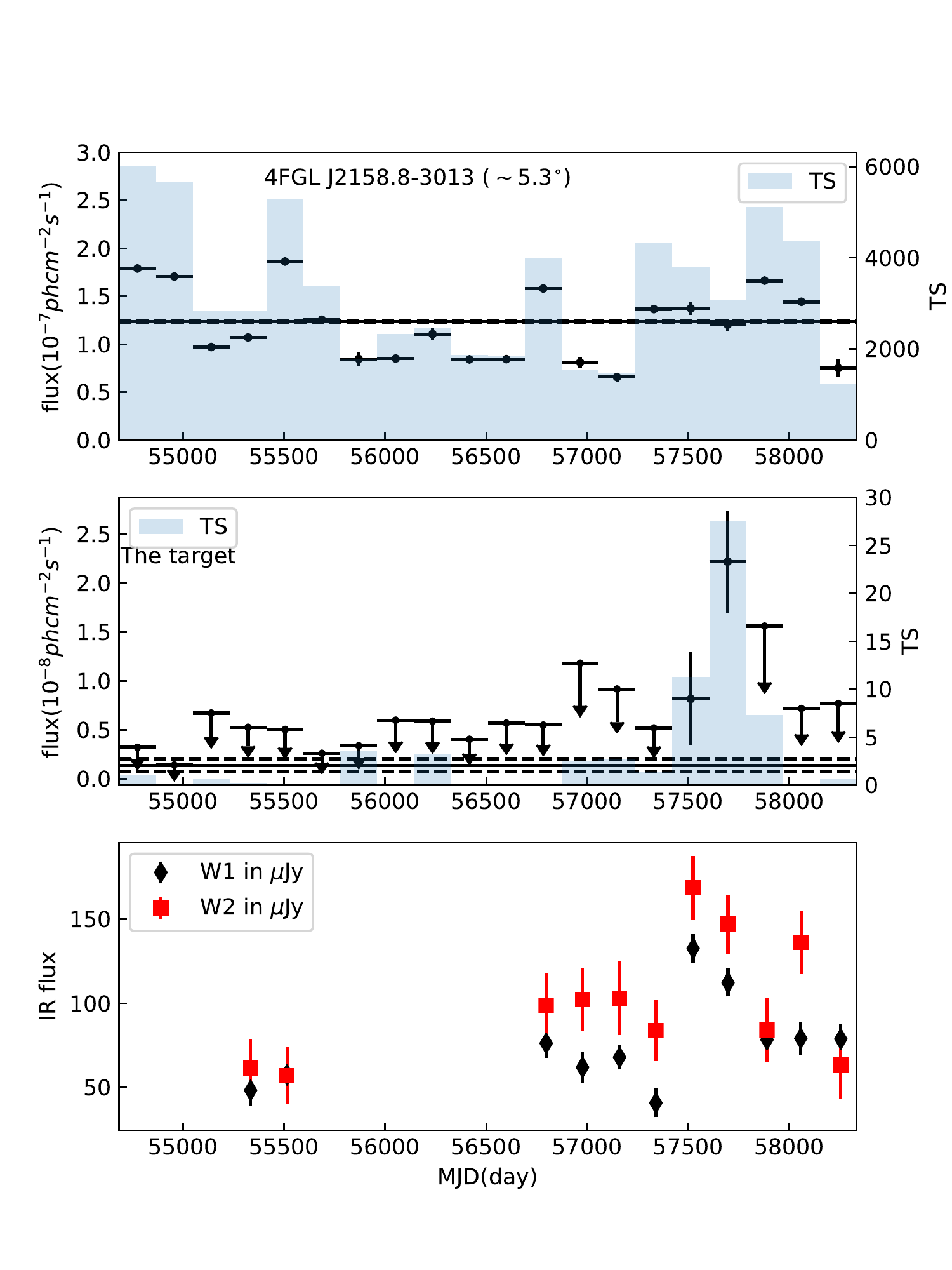}
  \label{fig:subfigure21}}
\subfigure[]{%
  \includegraphics[width=0.41\textwidth]{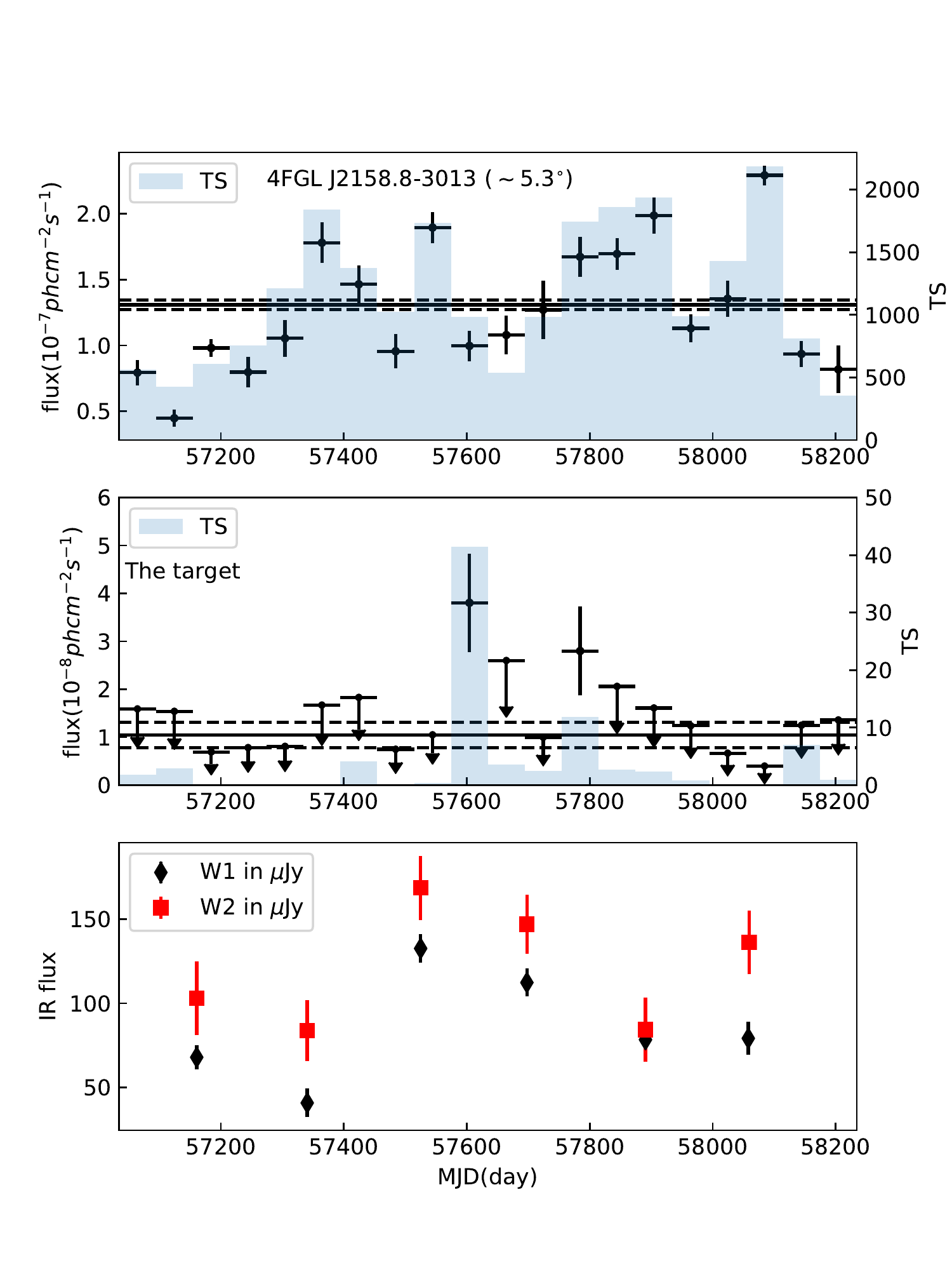}
  \label{fig:subfigure22}}
\caption{$\gamma$-ray and infrared light curves of PMN J2219-2719 as well as the $\gamma$-ray light curve of neighbor bright background source, (a) half-year bin, (b) 60-day bin. Horizontal solid line along with two dashed lines in each panel represent the average flux and its 1 $\sigma$ error with the whole time range, respectively.}  

\label{Fig.2}
\end{figure*}

The half-year bin light curve of PMN J2219-2719 displays only three consecutive bins (MJD 57422-57969, i.e., from 2016 February 4th to 2017 August 4th) with relatively large TS values (TS $>$ 4), including one bin with TS $>$ 25 that suggests an existence of a $\gamma$-ray transient (see the middle panel of Figure \ref{fig:subfigure21}). In order to quantify the significance of the variability, we apply the same method of \cite{2012ApJS..199...31N} to calculate the variability index and the corresponding variability significance level is 2.8 $\sigma$. Due to the limited angular resolution of $Fermi$-LAT, such phenomena could come from a nearby flaring bright background source \citep[e.g.,][]{2018ApJ...853..159L}. A very strong background source 4FGL J2158.8-3013 is ${5.33}^{\circ}$ away from the target. Considering the limited angular resolution of $Fermi$-LAT at 100MeV, we examine whether the new signal is contaminated by 4FGL J2158.8-3013. As shown in Figure \ref{fig:subfigure21}, the appearance of the new source is not coincident with any brightening of 4FGL J2158.8-3013. Furthermore, We extract the two-month bin $\gamma$-ray light curve of PMN J2219-2719 (see the middle panel of Figure \ref{fig:subfigure22}). In order to identify the exact flaring epoch, the monthly $\gamma$-ray light  curve is also extracted. During a period between 2016 July 5th and 2016 December 4th (MJD 57574-57726), a significant $\gamma$-ray signal (TS $\sim$ 45) appeared in the direction of PMN J2219-2719 (see Figure \ref{fig:subfigure12}), confirming our finding based on the monthly $\gamma$-ray light curve.

We carry out a localization analysis to identify the position of the new $\gamma$-ray source in the flare phase. The best fitted coordinates are R.A. ${334.85}^{\circ}$ and decl. ${-27.33}^{\circ}$ with a $2\sigma$ error radius of ${0.12}^{\circ}$. Besides, the angular distance between the new $\gamma$-ray position and the radio position of PMN J2219-2719 is only ${0.04}^{\circ}$ (see Figure \ref{fig:subfigure12}). The Bayesian association method (for details see \citealt{2010ApJS..188..405A}) was used to calculate the association probability. Our result shows the 96.9\% association probability and that is larger than the threshold of 80\%, which support the association between the $\gamma$-ray source and PMN J2219-2719. Adopting the new $\gamma$-ray position, a single power law function can well describe the $\gamma$-ray spectrum of the target. In the flare phase, the average photon flux reaches to ${(2.89 \pm 0.61) \times 10^{-8}~{\rm ph~cm^{-2} s^{-1}}}$, with a TS value of 47.8 ($\Gamma = 2.56 \pm 0.13$). If the high redshift FSRQ PMN J2219-2719 is the counterpart of the new $\gamma$-ray transient, the isotropic $\gamma$-ray luminosity is $(3.82 \pm 1.3) \times 10^{48}$ erg $\rm s^{-1}$. Due to the nearest background source is 2.4$^\circ$ away from the target and the 68\% C.L. contamination angles of LAT for 300~MeV photon is $\sim {2.3}^{\circ}$, we perform further analysis of the data with energy range from 300MeV to 500GeV, the TS value is then $\sim 35.5$. It is confirmed by the residual TS map, see Figure \ref{fig:subfigure13}. Following localization analysis suggests that PMN J2219-2719 still falls into the error radius of the $\gamma$-ray location.

\subsubsection{PMN J2321-0827}

\begin{figure}[ht!]
\centering
\subfigure[]{%
  \includegraphics[width=0.32\textwidth]{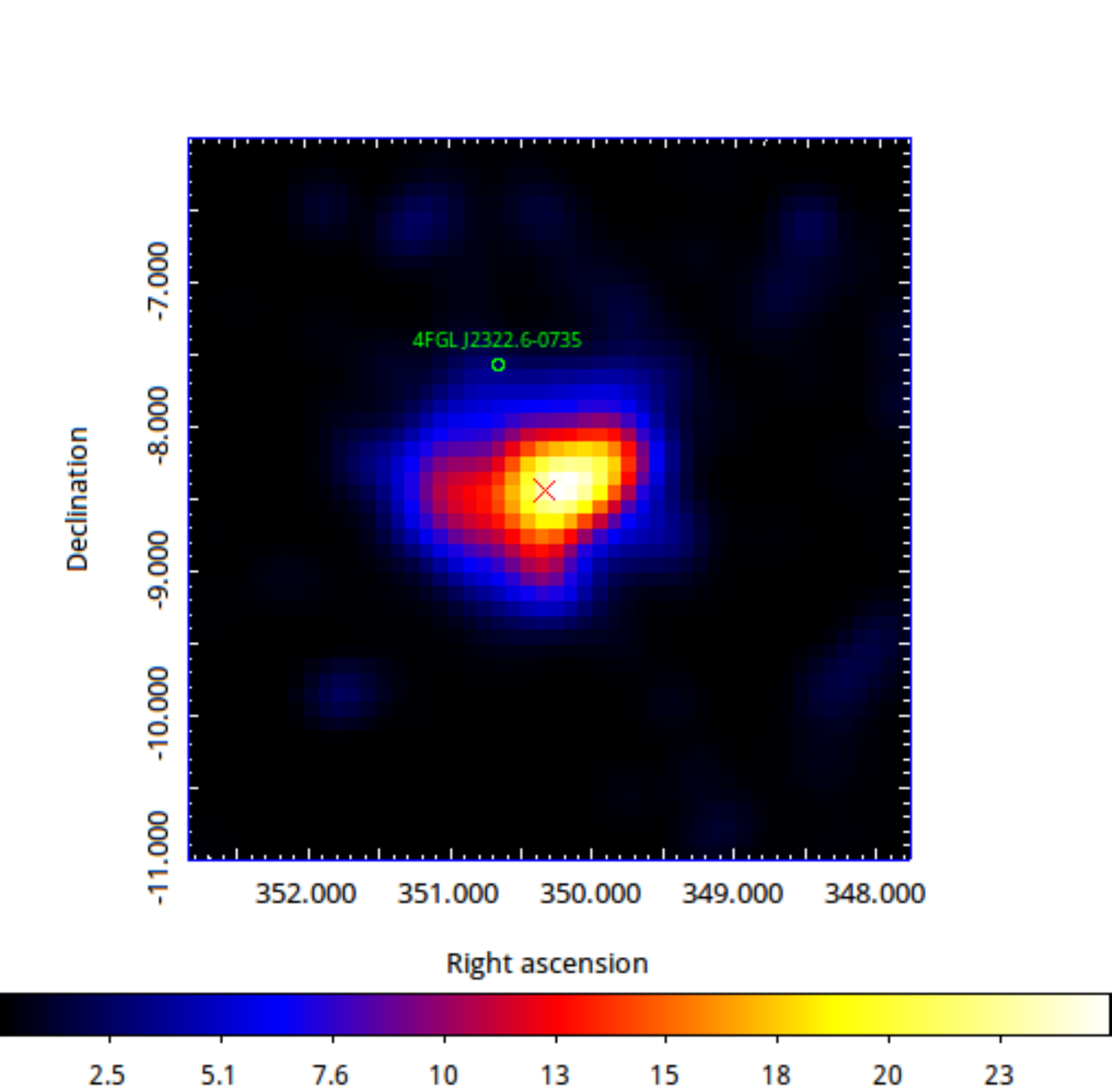}
  \label{fig:subfigure31}}
\subfigure[]{%
  \includegraphics[width=0.32\textwidth]{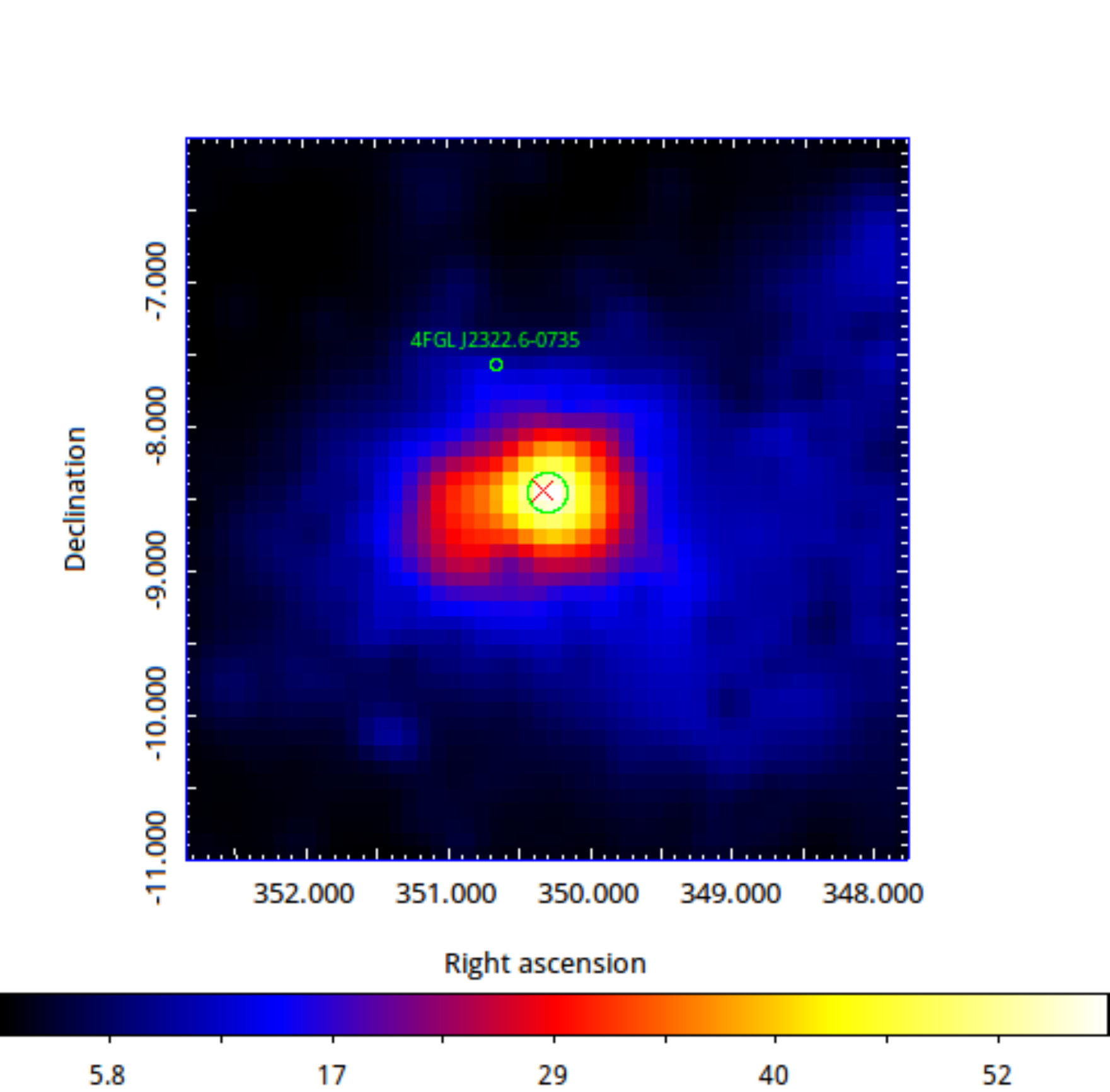}
  \label{fig:subfigure32}}
\subfigure[]{%
  \includegraphics[width=0.32\textwidth]{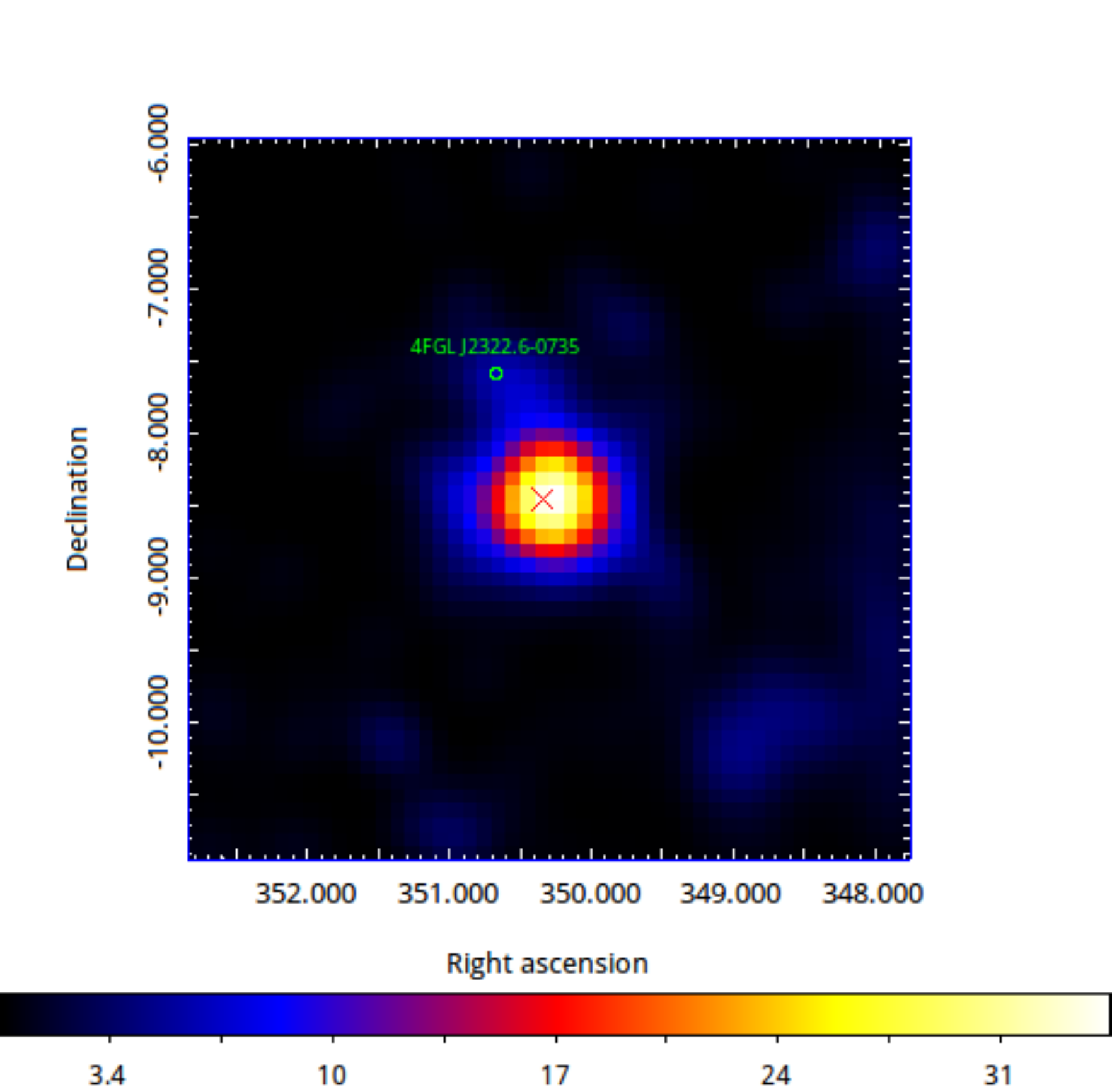}
  \label{fig:subfigure33}}
  \caption{TS maps for different epochs of PMN J2321-0827. The three panels represent $5^{\circ} \times 5^{\circ}$ TS maps with $0.1^{\circ}$ per pixel centered at PMN J2321-0827. (a) The 10-year TS map between 0.1 and 500 GeV. (b) 14-month flaring epoch TS map between 0.1 and 500 GeV. (c) TS map for the flaring epoch between 0.3 and 500 GeV (The 4FGL J2322.6-0735 has been removed from the model file.). The red X-shaped symbol represents the radio position (J2000) of PMN J2321-0827. In panel (b), the 95\% error circle of the $\gamma$-ray position is marked in green circle. Locations of the nearby background sources considered in the analysis are marked.}
\label{Fig.3}
\end{figure}

\begin{figure}[ht!]
\centering
\subfigure[]{%
  \includegraphics[width=0.41\textwidth]{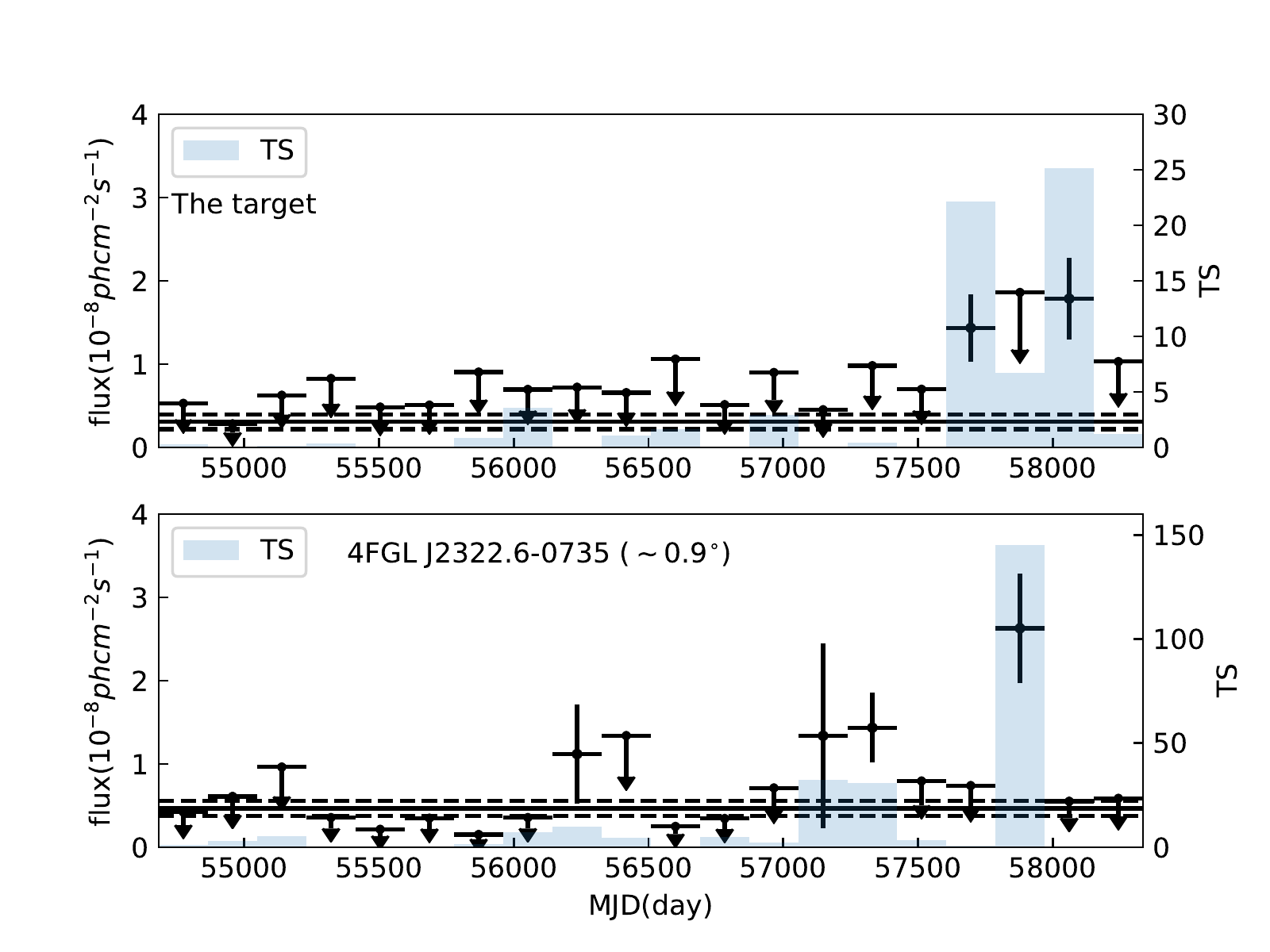}
  \label{fig:subfigure41}}
\subfigure[]{%
  \includegraphics[width=0.41\textwidth]{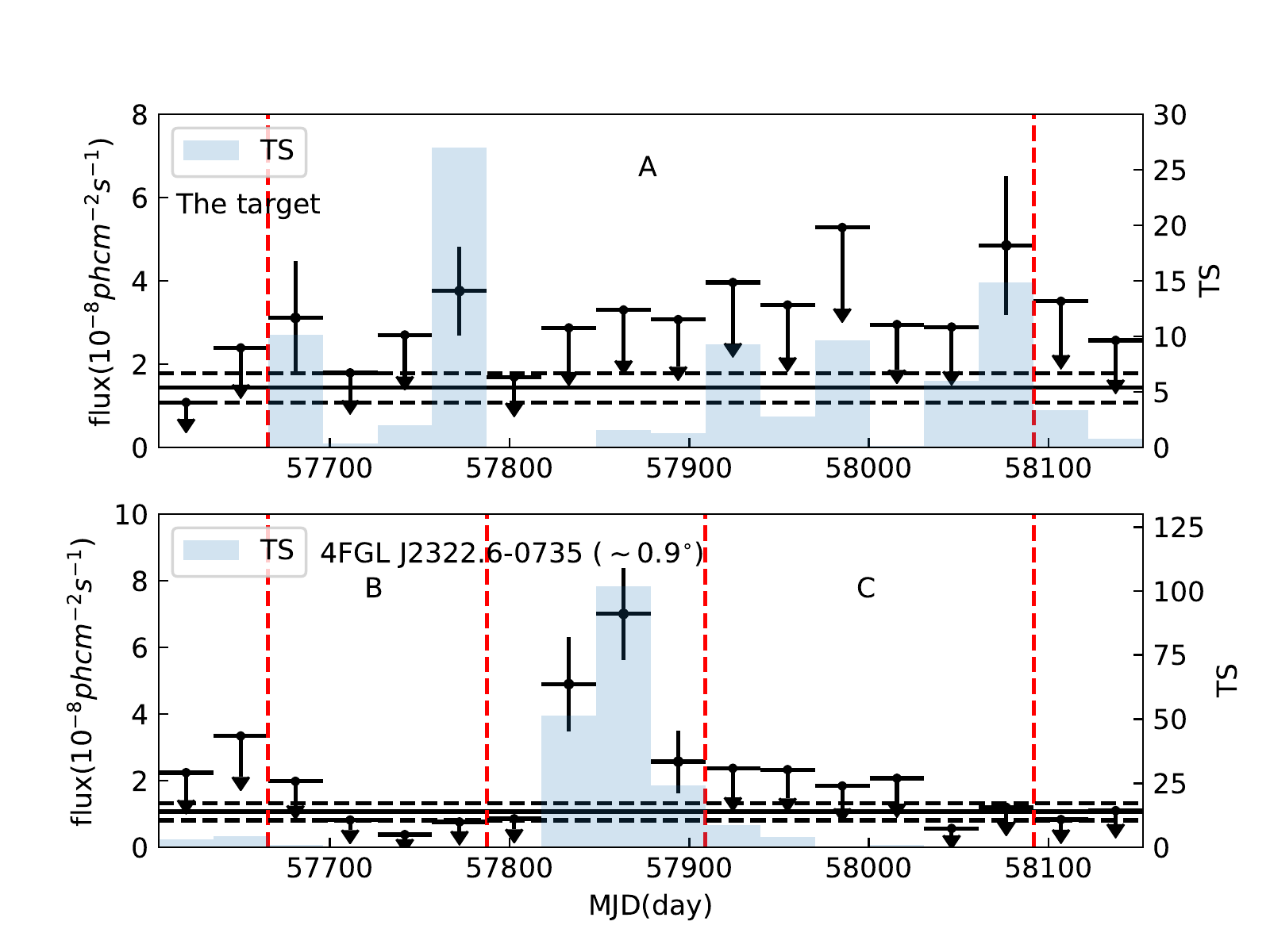}
  \label{fig:subfigure42}}
\caption{The light curves of PMN J2321-0827 and the neighbor background source. (a)half-year bin, (b) 30-day bin. Horizontal solid line along with two dashed lines in each panel represent the average flux and its 1 $\sigma$ error with the whole time range, respectively.}
\label{Fig.4}
\end{figure}

For the half-year bin $\gamma$-ray light curve, the TS values of the radiation in most bins are $<$ 4. However, in three consecutive bins (MJD 57604.7-58152.6, i.e., from  2016 August 4th to 2018 February 3th) the situations are different and in two bins we have TS $>$ 20 (see the upper panel of Figure \ref{fig:subfigure41}), which suggest the presence of a new transient source. The significance of $\gamma$-ray variability is 3.6 $\sigma$. One known $\gamma$-ray source in 4FGL (4FGL J2322.6-0735) is 0.9$^{\circ}$ away from the target. We also examined its temporal behaviors. As shown in Figure \ref{fig:subfigure41}, the temporal behaviors of the new $\gamma$-ray source is not coincident with its neighbor, which suggested that the flares are not contributed by 4FGL J2322.6-0735. Then we perform the fit in two time periods (MJD 54682-57604 and MJD 57604-58152, i.e., from 2008 August 4th to 2016 August 4th and 2016 August 4th to 2018 February 3th), respectively. No significant signal was found in the first period, which explains the absence of PMN J2321-0827 in 4FGL (since the chosen time period is consistent with that of the 4FGL catalog). However, a significant $\gamma$-ray signal (TS $\sim$ 55) appears at the location of PMN J2321-0827 in the second period and the result is consistent with the half-year bin $\gamma$-ray light curve. In addition, we extract the monthly light curve for the second period (see the upper panel of Figure \ref{fig:subfigure42}) to resolve the exact flare phase. A period of 14 months is marked by red dashed vertical line in the upper panel of Figure \ref{fig:subfigure42}, ranging from 2016 October 4th to 2017 December 4th (MJD 57665-58091). During this period, TS value of the $\gamma$-ray source towards PMN J2321-0827 is about 60 (see Figure \ref{fig:subfigure32}). The optimized $\gamma$-ray position of the new $\gamma$-ray source during the flare phase is R.A. ${350.29}^{\circ}$ and decl. ${-8.48}^{\circ}$, with a $2\sigma$ error radius of ${0.14}^{\circ}$. The $\gamma$-ray position is only offset ${0.05}^{\circ}$ from the radio position (see Figure \ref{fig:subfigure32}), with the association probability calculated as $94\%$. Our results suggested that PMN J2321-0827 is the counterpart of the new $\gamma$-ray source. Using the new $\gamma$-ray position, the average flux is estimated as ${(1.69 \pm 0.39) \times10^{-8}~\rm  ph~cm^{-2} s^{-1}}$, with a TS value of 61.4 and the spectral index is $\Gamma = 2.46 \pm 0.14$ and the isotropic $\gamma$-ray luminosity is $(1.49 \pm 0.38) \times 10^{48}$~erg $\rm s^{-1}$.

Further endeavors of eliminating the influence from 4FGL J2322.6-0735 are made. In addition to analyses of the entire flaring epoch (marked as A in the upper panel of Figure \ref{fig:subfigure42}), we also analyze the data in two separated ``periods" (marked as B, C in the lower panel of Figure \ref{fig:subfigure42}), when the TS values are 28.5 and 41.3, respectively. The radio position of PMN J2321-0827 is all within the $\gamma$-ray error circles. We then merge the data in the two periods, the joint analysis gives a TS value 67.4 and the corresponding $\gamma$-ray position is only offset by $0.03^{\circ}$ from the radio position. Note that in this case 4FGL J2322.6-0735 is removed from the background model file due to its small TS value ($<$ 5), and the nearest background source is $\sim$ 2.3$^{\circ}$ away. Since the 68\% C.L. contamination angles of LAT for 300~MeV and 1~GeV photon decrease to $\sim {2.3}^{\circ}$ and $\sim {0.8}^{\circ}$, individual analyses of {\it Fermi}-LAT data between 300~MeV and 500~GeV as well as between 1~GeV and 500~GeV are performed, and the TS values are 46.5 and 33, respectively. The results are confirmed by the corresponding residual TS Maps, see Figure \ref{fig:subfigure33}. In conclusion, the $\gamma$-ray source is robust and likely associated with PMN J2321-0827.

\section{Discussion} \label{sec:diss}

So far only five FSRQs with redshift greater than 3.1 have been detected by $Fermi$-LAT in the latest catalog of AGN (4LAC) \citep{2020ApJ...892..105A}, three of them with redshift larger than 3.5. Our results enrich the samples of $\gamma$-ray sources with redshift $>3.1$. Particularly, PMN J2219-2719 (z=3.63) is the fourth farthest $\gamma$-ray FSRQ. The high redshift $\gamma$-ray sources carry the information of the extragalactic background light (EBL). However, the most energetic $\gamma$-ray photons from the transients of PMN J2219-2719 and PMN J2321-0827 are just 10.8 GeV and 5.1 GeV, respectively, which are insufficient to challenge the current EBL models ($ E_{\rm horizon}\sim40~\rm GeV$ for z $\sim 3$, \citealt{2010ApJ...712..238F}). Detections of high redshift $\gamma$-ray FSRQs are helpful to reconstruct the evolution of the $\gamma$-ray luminosity function (GLF) of blazars at high redshift. Since so far the most distant $\gamma$-ray sample to calculate the blazar GLFs \citep[e.g.,][]{2012ApJ...751..108A,2013MNRAS.431..997Z} is $z$ = 3.1. Therefore, the detection of new high redshift $\gamma$-ray FSRQs, especially the new sources with redshift $>3.1$ is very important to update the GLF of blazars. 

\begin{figure}[ht!]
\centering
\includegraphics[width=0.7\textwidth]{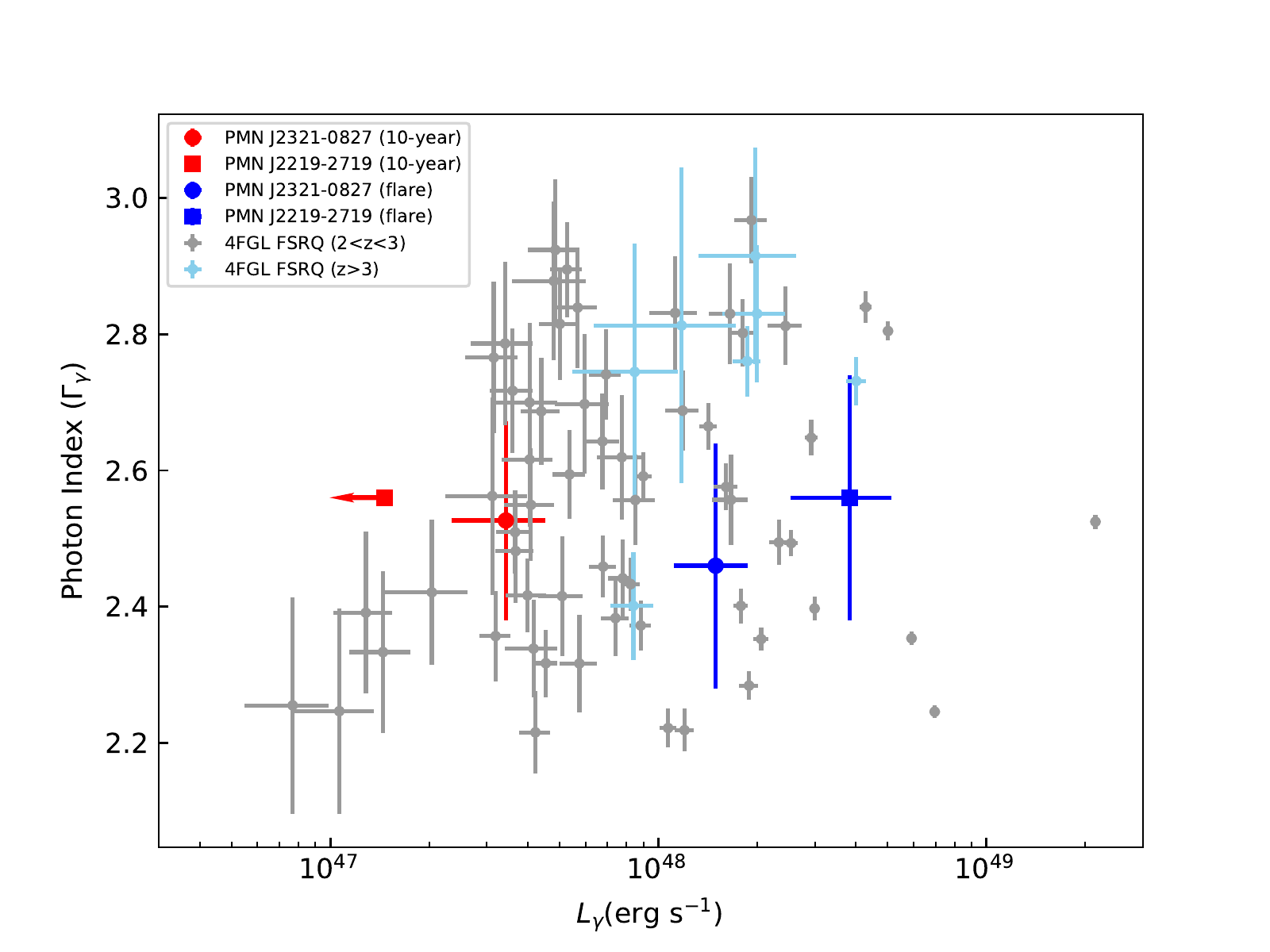}
\caption{Comparison of the two newly identified high redshift $\gamma$-ray FSRQs with the 4FGL sources: $\gamma$-ray luminosity vs. photon index. The plotted $L_{\gamma}$ and $\Gamma_{\gamma}$ are derived from the energy range between 100MeV and 300GeV. The time-averaged photon index of PMN J2219-2719 is fixed to 2.56.}
\label{Fig.6}
\end{figure}

In Figure \ref{Fig.6}, we compare the two newly detected distant objects with the high redshift FSRQs (z $>$ 2) reported in the 4FGL catalog. The 10-year averaged emission of these two sources locate in the region of low $\gamma$-ray luminosities ($L_{\gamma} <  5 \times 10^{47}$~erg $\rm s^{-1}$). While in the flare phases, the $\gamma$-ray luminosities are similar to that of high redshift FSRQs. This is not surprising since we can only detect the distant sources with a high luminosity. Likely there are much more dim $\gamma$-ray FSRQs, which can only be detected when they are flaring or alternatively by a future space GeV detector that is much more sensitive than $Fermi$-LAT.

Generally, simultaneous multi-wavelength observations are crucial to pin down the counterpart. So far, simultaneous $\gamma$-ray and optical flares have been frequently detected for FSRQs \citep[e.g.,][]{2010Natur.463..919A}. In addition, simultaneous $\gamma$-ray and infrared flare also have been detected for the high redshift blazar CGRaBSJ0733+0456 \citep{2019ApJ...879L...9L}. Recently, the simultaneous brightening of $\gamma$-ray and optical emissions of a high-redshift blazar GB 1508+5714 is reported in \cite{Liao:2020bek}.
Therefore, the simultaneous multi-wavelength observations for PMN J2219-2719 and PMN J2321-0827 are very important. Fortunately, for PMN J2219-2719, a sign of WISE W1 and W2 flare appeared between 2015 and 2016, overlaps with the $\gamma$-ray flare (see Figure \ref{fig:subfigure21}), which suggest that PMN2219-2719 is the low energy counterpart of the $\gamma$-ray source. On the other hand, due to the limited data, it is not possible to directly establish the association of the $\gamma$-ray and the infrared emission in the 60-day bin light curves (see Figure \ref{fig:subfigure22}). Meanwhile, we have tried to find other potential counterparts, especially blazars or blazar candidates included in the BZCAT list \citep{2009A&A...495..691M} and other radio surveys \citep[e.g.,][]{1998AJ....115.1693C,2003MNRAS.341....1M,2007ApJS..171...61H,2008ApJS..175...97H}. For PMN J2219-2719, we do not find other potential counterparts within the $2\sigma$ error radius. These facts support PMN J2219-2719 is the counterpart of the $\gamma$-ray source.
A transient $\gamma$-ray source towards B3 1428+422 (z = 4.72) is detected by $Fermi$-LAT \citep{2018ApJ...865L..17L}, no simultaneous multi-wavelength data can be used to pin down the relationship between the transient $\gamma$-ray source and B3 1428+422. Similarly, we do not find the simultaneous multi-wavelength observations for PMN J2321-0827 and can not determine the low energy counterpart. The association probability and the spatial association between the $\gamma$-ray source and PMN J2318-0827 suggest that PMN J2318-0827 is the counterpart of the $\gamma$-ray source. According to the $\gamma$-ray position information, another potential counterpart locates in the 95\% C.L. $\gamma$-ray error radius, which is a radio loud (RL = 119) narrow-line Seyfert 1 (NLSY1) FBQS J2321-0825 (z=0.45), as reported in \cite{2017A&A...603A.100L}. The radio position of FBQS J2321-0825 is offset $0.06^{\circ}$ from the best fit $\gamma$-ray position. The association probability is $56.2\%$ under the threshold of 80.0\%. Generally, flat spectrum radio source tends to produce $\gamma$-ray emission. However, the radio emission of FBQS J2321-0825 has been detected in only one frequency \citep[S$_{\rm 1.4GHz} = 5.0$ mJy,][]{2017A&A...603A.100L} so far, and we don't know the radio spectrum information. Furthermore, the RL parameter of the known $\gamma$-ray NLSY1 (RL $>$ 250) in 4FGL are much more than FBQS J2321-0825 (RL = 119). In addition, PKS 1502+036 \citep[$z=0.41$,][]{2015ApJS..219...12A} is a known $\gamma$-ray NLSY1 in 4FGL, which radio luminosity at 1.4 GHz ($L_{\rm1.4GHz}$ $\sim 3.3 \times 10^{42}$~erg $\rm s^{-1}$) is about 58 times of the FBQS J2321-0825 ($L_{\rm1.4GHz}$ $\sim 5.6 \times 10^{40}$~erg $\rm s^{-1}$). Motivated by these facts, we suggest that PMN J2321-0827 also has the low energy counterpart. The future simultaneous multi-frequency observations in the period of the high $\gamma$-ray activity of the new transient can be used to confirm association with the lower-energy counterpart.

\begin{figure}[ht!]
\centering

  \includegraphics[width=0.7\textwidth]{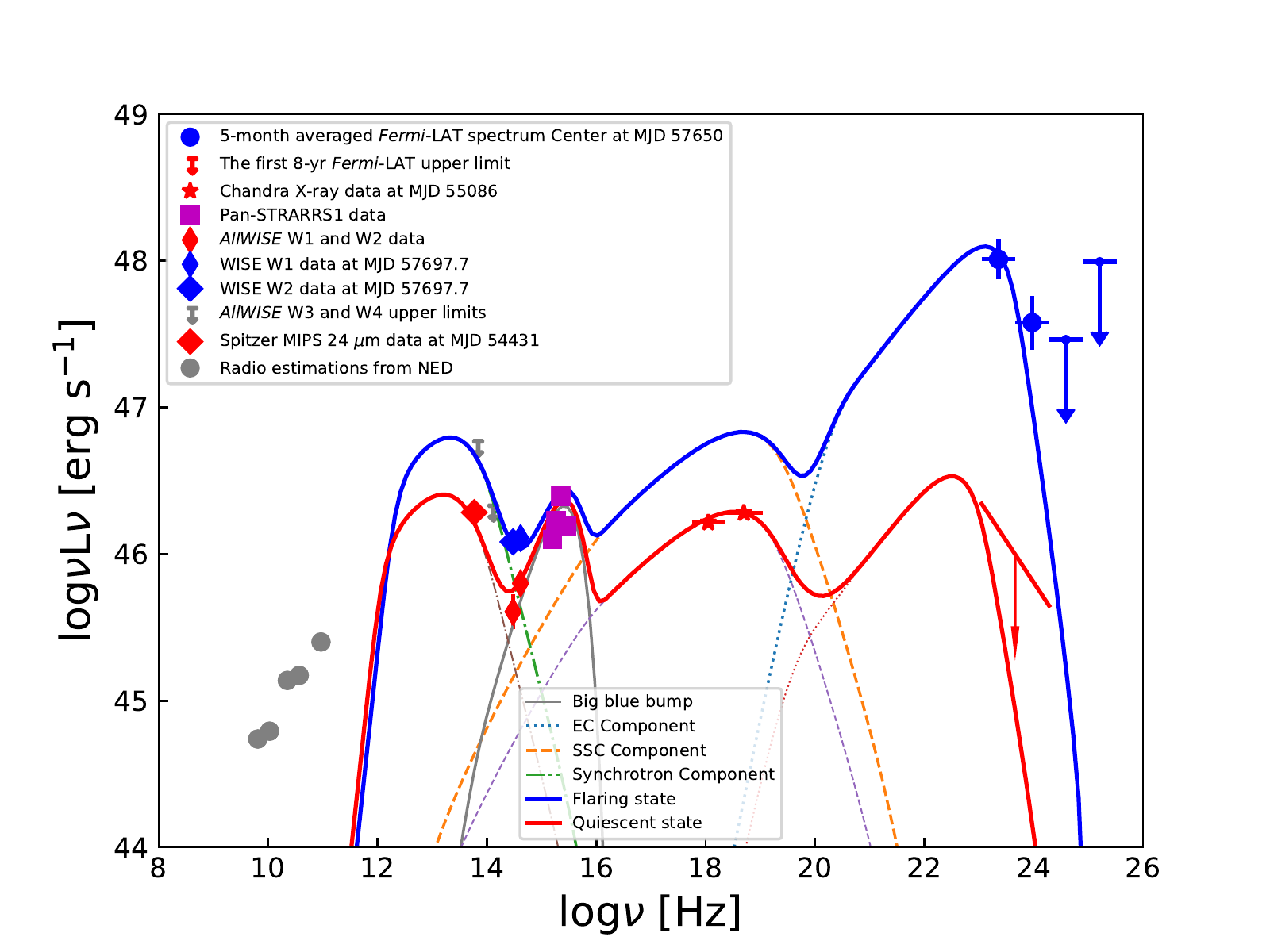}
\caption{The broadband Spectral Energy Distributions (SEDs) in different flux states of PMN 2219-2719 along with the theoretical descriptions. The color blue and red lines correspond to the flaring and the quiescent states, respectively.}
\label{Fig.7}
\end{figure}

Broadband SEDs of PMN J2219-2719 are shown in Figure \ref{Fig.7}. Multi-wavelength data including radio flux densities obtained from NED, median five-band optical fluxes from the second data release from the Panoramic Survey Telescope and Rapid Response System (Pan-STARRS, \citealt{2016arXiv161205560C,2016arXiv161205243F}) are collected. Meanwhile, the X-ray fluxes from {\it Chandra} observation on 12th September 2009 \citep{2011ApJ...738...53S}, ALLWISE W1 and W2 fluxes and Spitzer MIPS 24 $\mu$m flux observed on 27th November 2007, and the first 8-yr {\it Fermi}-LAT 95 C.L. upper limit represent the low flux state SED, while WISE W1 and W2 fluxes on 5th November 2016 and a 5-month averaged {\it Fermi}-LAT $\gamma$-ray spectrum centered on 19th September 2016 correspond to the high flux state SED. The Pan-STARRS and the ALLWISE data are extracted by the emission from a standard \cite{1973A&A....24..337S} disk, extending from 3$R_{s}$ to 2000$R_{s}$, where $R_{s}$ is the Schwarzschild radius. The accretion disk produces a total luminosity $L_{d}=\eta\dot{M}c^{2}$ in which $\dot{M}$ is the accretion rate and the accretion efficiency $\eta$ is set as a typical value, 0.1. The accretion disk emission follows a multi-temperature radial profile, and the local temperature at a certain radius $R_{d}$ is,
\begin{equation}
T^{4} = \frac{3R_{s}L_{d}}{16\pi\eta\sigma_{MB}R_{d}^{3}}\left[1-(\frac{3R_{s}}{R_{d}})^{1/2}\right].
\end{equation}   
The nonthermal jet emission is described by a simple homogeneous one-zone leptonic scenario, including synchrotron and IC (both SSC and EC) processes along with the synchrotron self-absorption process and the Klein$-$Nishina effect in the IC scattering. A relativistic compact blob with a radius of $R_{j}^{\prime}$ embedded in the magnetic field is responsible for the jet emission, with an assumption that emitting electrons follow a broken power-law distribution, 
\begin{equation}
N(\gamma )=\left\{ \begin{array}{ll}
                    K\gamma ^{-p_1}  &  \mbox{ $\gamma_{\rm min}\leq \gamma \leq \gamma_{br}$} \\
            K\gamma _{\rm br}^{p_2-p_1} \gamma ^{-p_2}  &  \mbox{ $\gamma _{\rm br}<\gamma\leq\gamma_{\rm max}$,}
           \end{array}
       \right.
\label{Ngamma}
\end{equation}
where the $p_{1,2}$ are indices of the broken power-law particle distribution, $\rm \gamma_{br}$ is the electron break energy, $\rm \gamma _{min}$ and $\rm \gamma _{max}$ are the minimum and maximum energies of the electrons, and $K$ is the normalization of the particle number density. Meanwhile, the radius of the emitting blob is often constrained by the variability timescale, $R_{j}^{\prime} \le ct_{var}\delta(1+z)^{-1}$, where c is the speed of the light and $\delta$ is the Doppler factor of the jet blob. Due to the limited statistic, such information is not available for PMN J2219-2719. Nevertheless, we set typical values of the variability timescale, 1 day corresponding to the high flux state SED and 5-day for the low flux state SED. In addition, the Ly$\alpha$ line emission is adopted as the external soft photons in the EC process, with the energy density set as $2\times10^{-2}$ erg $\rm cm^{-3}$ \citep{2012MNRAS.425.1371G}. The transformations of frequency and luminosity between the jet frame and the observational frame are $\nu = \delta\nu^{\prime}/(1+z)$ and $\nu L_{\nu} = \delta^{4}\nu^{\prime}L^{\prime}_{\nu^{\prime}}$. The simple leptonic model can give reasonable descriptions of both SEDs (see Figure \ref{Fig.7}). The input parameters of the jet radiation models are listed in Table \ref{tpara}. A possible ejecta of a new jet blob that leads a significant enhance of the Doppler factor could account for the $\gamma$-ray and optical brightening of PMN J2219-2719. The SEDs of high redshift blazars have been analyzed \citep{2011MNRAS.411..901G, 2020ApJ...889..164M}. Their typical magnetic field strength is $\sim$ 1 Gauss and bulk Lorentz factor of the jet blob is $\sim$ 13. Recently, \cite{2020arXiv200601857P} perform a systematical SED modeling study of 142 high redshift ($z >$ 3) blazars. The average magnetic field strength is $\langle B \rangle$ = 1.0 Gauss and Doppler factor of the jet blob is  $\langle \delta \rangle$ = 12.3. For flaring state of PMN J2219-2719, the parameters of the jet radiation models ($B$ and $\delta$) are consistent  with the results of \cite{2020arXiv200601857P}. QSO J0906+6930 is the farthest known blazar and it has not been identified as a $\gamma$-ray source so far,  which SED is revisited  \citep{2018ApJ...856..105A}. Moreover, the parameters of radio jet of QSO J0906+6930 are determined based on VLBI data: Doppler factor = 6.1 $\pm$ 0.8 and Lorentz factor =3.6 $\pm$ 0.5 \citep{2020NatCo..11..143A}. In order to explain the $\gamma$-ray ``off"  SED, the bound is set as 6$< \delta <$ 11.5. In short, the input parameters of the jet radiation models here are consistent with SED modeling studies for other high redshift blazars.

\section{Conclusion}

In this work, we have systematically analyzed the $Fermi$-LAT Pass 8 data of two high-redshift FSRQs (PMN J2219-2719 and PMN J2321-0827). In the following, we give a brief summary.
\begin{enumerate}

\item We report the first time detection of  $\gamma$-ray signals in the direction of PMN J2219-2719 and PMN J2321-0827. Our findings enrich the samples of $\gamma$-ray sources with redshift $>3$.

\item During the flare period of PMN J2219-2719, the flux is ${2.89 \times10^{-8}~ {\rm ph~cm^{-2}~ s^{-1}}}$($\Gamma$= $2.56 \pm 0.13$), the corresponding TS value is 47.8. Comparing with the 95\% c.l. upper limit of 10-year averaged flux (${2.58 \times10^{-9}~{\rm ph~cm^{-2}~s^{-1}}}$), the $\gamma$-ray flux enhancement is by a factor of 10. In the flare phase of PMN J2321-0827, the TS value is 61.4 and the flux is ${1.69 \times 10^{-8}~ \rm ph~cm^{-2} s^{-1}}$, about 4 times of  the 10-year averaged flux (${3.8\times10^{-9}~{\rm ph~ cm^{-2} s^{-1}}}$). Likely there are much more dim high redshift $\gamma$-ray FSRQs, which can only be detected when they are flaring.

\item Considering the similarity between the $\gamma$-ray and infrared light curves and the spatial association between the $\gamma$-ray source and PMN J2219-2719, we suggest that PMN J2219-2719 is the low energy counterpart of the new transient GeV source.

\item The broadband SED of PMN J2219-2719, during the high activity state, is similar to that observed from other high redshift blazars. 

\end{enumerate}

In the future, the number of high-redshift $\gamma$-ray sources will increase due to the continual successful operation of $Fermi$-LAT. Moreover, with upcoming wide-deep-fast sky survey facilities, such as the Large Synoptic Survey Telescope \citep{Ivezic:2008fe} and other future observational facilities in time domain (e.g. the Wide-Field InfraRed Survey Telescope, \citealt{2012arXiv1208.4012G}; the Einstein Probe, \citealt{Yuan:2015tia}), a comprehensive broadband dynamic view of high redshift $\gamma$-ray sources will be achieved.

\section*{acknowledgments}

We appreciate the helpful suggestions from the anonymous referee. This work was supported in part by the NSFC under grants 11525313 (i.e., Funds for Distinguished Young Scholars) and 11703093.

This work has made use of data obtained from the High Energy Astrophysics Science Archive Research Center (HEASARC), provided by NASA¡äs Goddard Space Flight Center. This work also makes use of the SIMBAD database, operated at CDS, Strasbourg, France. This work has made use of the NASA/IPAC Extragalactic Database (NED), which is operated by the Jet Propulsion Laboratory, California Institute of Technology, under contract with the National Aeronautics and Space Administration. This work makes use of data products from the Wide-field Infrared Survey Explorer, which is a joint project of the University of California, Los Angeles, and the Jet Propulsion Laboratory/California Institute of Technology, funded by NASA. This work has also used IPython \citep{2007CSE.....9c..21P}, NumPy, SciPy \citep{2011CSE....13b..22V}, Matplotlib \citep{2007CSE.....9...90H}.

\facilities{$Fermi$-LAT, WISE}

\bibliographystyle{aasjournal}

\begin{deluxetable}{lccccccccc}[h]
\scriptsize
\tablenum{1} \tablewidth{0pt}
\tablecaption{List of parameters used to construct the theoretical jet SEDs in Figure \ref{Fig.7}.}
\tablehead{ \colhead{Model} &\colhead{$p_{1}$} &\colhead{$p_{2}$} &\colhead{$\gamma_{br}$} &\colhead{$K$[$\rm cm^{-3}$]} &\colhead{$B$[Gauss]} &\colhead{$\delta$} &\colhead{$R_{j}^{\prime}$[cm]}}
\startdata
Flaring state &2.0 &6.0 &496 &$\rm 3.5\times10^{5}$ &2.1 &16.7 &$\rm 9.3\times10^{15}$ \\[3pt]
Quiescent state &2.0 &6.0 &476 &$\rm 1.2\times10^{5}$ &4.1 &7.2 &$\rm 2.0\times10^{16}$ \\[3pt]
\enddata
\tablecomments{$p_{1,2}$ are the indexes of the broken power-law radiative electron distribution; $\gamma_{br}$ is the break energy of the electron distribution; $K$ is the normalization of the particle number density; $B$ is the magnetic field strength; $\delta$ is the Doppler boosting factor and $R_{j}^{\prime}$ is the radius of the emission blob in the jet comoving frame. The minimum and maximum energies of the electrons are set as 100 and 10 times of the $\gamma_{br}$, respectively. The energy density of the Ly$\alpha$ line emission is estimated as $2\times10^{-2}$ erg $\rm cm^{-3}$ in the rest frame. \tiny} 
\label{tpara}
\end{deluxetable}

\end{document}